\documentstyle[12pt,epsf]{article}

\newcommand{\be}{\begin{equation}}
\newcommand{\ee}{\end{equation}}
\newcommand{\eel}[1]{\label{#1}\end{equation}}
\newcommand{\bea}{\begin{eqnarray}}
\newcommand{\eea}{\end{eqnarray}}
\newcommand{\eeal}[1]{\label{#1}\end{eqnarray}}
\newcommand{\baq}{\begin{equation}\begin{array}{rcl}}
\newcommand{\eaq}{\end{array}\end{equation}}
\newcommand{\eaql}[1]{\end{array}\label{#1}\end{equation}}
\newcommand{\beac}{\begin{equation}\begin{array}{rcl}}
\newcommand{\eeacn}[1]{\end{array}\label{#1}\end{equation}}
\newcommand{\ba}{\begin{array}}
\newcommand{\ea}{\end{array}}

\newcommand{\equ}[1]{(\ref{#1})}

\renewcommand{\a}{\alpha}

\newcommand{\RR}{{\cal R}}

\newcommand{\ns}{NS$^\prime~$}
\newcommand{\nss}{NS$^{\prime\prime}~$}

\newcommand{\notR}{\not{\hbox{\kern-4pt $R$}}}
\newcommand{\journal}[4]{{\rm #1~}{{\rm #2}}\,(19#3)\,#4}

\newcommand{\np}{\journal {Nucl. Phys.}}
\newcommand{\pl}{\journal {Phys. Lett.}}

\newcommand{\PRD}{{\em Phys. Rev. D}}
\begin{document}

\begin{titlepage}

{\sf
\begin{flushright}
{TAUP--2415--97, LMU--TPW--97--7 }\\
{hep--th/9704044}\\
{April 1997}
\end{flushright}}
\vfill
\vspace{-1cm}
\begin{center}
{\large \bf Brane Configurations and 4D Field Theory Dualities}
\vskip 1cm
{\sc A. Brandhuber$^{a,}$\footnote{Work supported
    in part by the US-Israel Binational Science Foundation, by GIF --
    the German-Israeli Foundation for Scientific Research, and by the
    Israel Science Foundation.}, J. Sonnenschein$^{a,1}$,
    S. Theisen$^{b,}$\footnote{
    Work supported   in part by 
    GIF --
    the German-Israeli Foundation for Scientific Research  
 and by the European Commission TMR programme ERBFMRX-CT96-0045,
    in which S.T. is associated to HU-Berlin.}
 and S. Yankielowicz$^{a,1}$,}\\[10mm]
{\em $^a$School of Physics and Astronomy,}\\
{\em Beverly and Raymond-Sackler Faculty of Exact Sciences,}\\
{\em Tel-Aviv University, Ramat-Aviv, Tel-Aviv 69978, Israel}\\[5mm]
{\em and}\\[5mm]
{\em $^b$Sektion Physik, Universit\"at M\"unchen, }\\
{\em Theresienstra\ss e 37, 80333 M\"unchen, FRG}

\end{center}
\vfill

\thispagestyle{empty}

\begin{abstract}

We study brane configurations which correspond to field theories in
four dimension with N=2 and  N=1 supersymmetry. In particular we
discuss brane motions that translate to Seiberg's duality in N=1
models recently studied by Elitzur, Giveon and Kutasov. We
investigate, using the brane picture, the moduli spaces of the dual
theories. Deformations of these models like mass terms and vacuum
expectation values of scalar fields can be identified with positions
of branes. The map of these deformations between the electric and dual
magnetic theories is clarified. The models we study reproduce known
field theory results and we provide an example of new dual pairs with
N=1 supersymmetry. Possible relations between brane configurations and
non-supersymmetric field theories are discussed.

\end{abstract}

\end{titlepage}

\section{Introduction}

Recently, it has become clear that many of the dualities and
exact results in supersymmetric
field theories have direct realizations in terms of
string theory. The main idea is to construct a local description of
the field theory in a string theory setup. This local description
involves geometric aspects of the compactification manifold
\cite{kklmv,kkv}, the local geometry together with D-branes wrapped around
homology cycles \cite{bjpsv}.
The simplest approach, however, is the one taken in \cite{han}, where
one finds an arrangement of {\it flat}
D and NS branes in $d=1$ Minkowski space,
such that the field theory on the world-volume of the D-branes
has the desired gauge symmetry, matter content and
number of supersymmetries.

In \cite{han} Dirichlet threebranes ending on Dirichlet fivebranes or
solitonic fivebranes in IIB string theory were used to study field
theories in three dimensions. The fivebranes are heavier than the
threebranes and can be treated classically whereas the world-volume of the
threebrane is finite in one direction. Thus, at low energies the
threebrane worldvolume appears to be three dimensional. The
world-volume theory of threebranes stretched between two solitionic
branes has $N=4$ supersymmetry in three dimensions and the inclusion
of Dirichlet fivebranes leads to matter multiplets in the world-volume
theory coming from open strings between the Dirichlet fivebranes and
the Dirichlet threebranes. Brane
configurations corresponding to field theories in three dimensions with
$N=4$ were also studied in \cite{boer1,boer2}. Threebranes stretched
between two different types of solitonic fivebranes which lead to
field theories with $N=2$ supersymmetry were investigated in \cite{boer1}.

A rather similar construction in IIA string theory was introduced by
Elitzur, Giveon and Kutasov \cite{egk} to analyse four dimensional
theories with $U(n)$ gauge groups.
The generalization to $SO(N)$ and $USp(N)$ gauge groups
can be found in \cite{ejs}.
In this case
Dirichlet fourbranes stretched between two types of solitonic
fivebranes lead, at low energy, to an effective {\sl four} dimensional
world-volume field theory with $N=1$ supersymmetry since
one of the fourbrane directions is finite. The addition of Dirichlet
sixbranes (replacing the Dirichlet fivebranes in IIB) corresponds to
additional chiral matter multiplets.
By making certain deformations in
the brane configuration Seiberg's duality \cite{sei1} and the duality
of theories with adjoint matter \cite{kut} can be realized.

The aim of this paper is to
study in general brane configurations in IIA which correspond to field
theories in four dimensions with $N=2$, $N=1$ and possibly $N=0$ supersymmetry.
In particular using brane rearrangements we provide further
evidence to

(i) the equivalence 
between the Higgs branches of
``dual" $N=2$ models,

(ii) the correspondence between the deformations
of the $N=1$ electric and magnetic theories associated with Seiberg's
duality,

(iii) generalization of the latter theories to theories
with product gauge groups.

In section 2 we explain the basic types of brane configurations in
type IIA string theory used in this paper and treat other general
aspects as the classical
global symmetries of the world-volume theories and their partial
breaking by instantons, and creation/annihilation processes of fourbranes.
In section 3, we
analyse models with $N=2$ supersymmetry and provide a brane
description of the duality in the Higgs branches of these theories.
In section 4, we study $N=1$ supersymmetric theories. We review the brane
description of Seiberg's duality and adjoint duality \cite{egk} and precisely
identify deformations of the theories and work out the map between the
deformations in the electric theories and the dual magnetic
theories. Furthermore we study a configuration with three solitonic
fivebranes which provides a new dual pair of theories with product
gauge groups.
In section 5, we try to make contact with dualities in gauge theories
without any supersymmetry. We introduce a set of branes that breaks all 32
supersymmetries of IIA.  We analyze the possibility for having brane
configurations which may be related to duality in non-supersymmetric
QCD with adjoint fermions.
In section 6, we address the problem of identifying field theory
instantons in the brane setup.

\section{From brane configurations to field theory - generalities}

We will use the same types of branes in type IIA string theory which
were
introduced in \cite{egk}: two types of Neveu-Schwarz five branes
NS and NS$^\prime$ whose world-volumes have coordinates $x^0, x^1, x^2,
x^3, x^4, x^5$ and $x^0, x^1, x^2, x^3, x^8, x^9$, respectively, a Dirichlet
sixbrane D6 whose world-volume has coordinates $x^0, x^1, x^2, x^3,
x^7, x^8, x^9$ and a Dirichlet fourbrane D4 with world-volume $x^0,
x^1, x^2, x^3, x^6$. This set of branes preserves 1/8 of the 32
supercharges of the type IIA string theory. A T-duality transformation
in the $x^3$ direction reproduces the configuration in IIB theory
recently studied in \cite{boer1}. Note that configurations without the
NS$^\prime$ type of branes break only 3/4 of the supersymmetry of the original
theory and T-duality in the $x^3$ direction gives the configurations
studied in \cite{han} which correspond to 3d gauge field theories with
N=4 supersymmetry.\\

\begin{figure}[ht]
\hbox to\hsize{\hss
\epsfysize=4cm
\epsffile{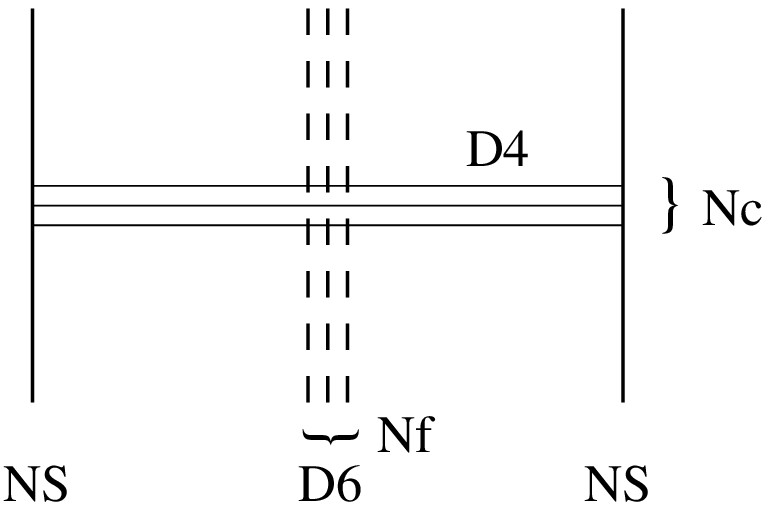} \hss}
\vspace{0cm}
\caption{} \label{pic1a}
\end{figure}

\underline{$N=2$ configurations} (cf. figure \ref{pic1a})\\

We start with the simplest configuration involving two NS branes
at equal position in the $x^7, x^8, x^9$ directions and at
different locations in the $x^6$ direction. Between
the two NS branes $N_c$ D4 branes are stretched. Their location in
the $x^7,x^8,x^9$ directions is fixed and their world-volume is
${\bf R}^{1,3}$ times a finite interval in the $x^6_{{\rm NS}} < x^6 <
x^6_{{\rm NS}^\prime}$ direction. The world-volume theory on the D4 branes at
long distances (compared to the $x^6$ direction) is a
four dimensional $N=2$
supersymmetric Yang-Mills theory with gauge group $U(N_c)$. The $N=2$
vector multiplet consists of a $N=1$ vector multiplet $W_\alpha$ which
is associated with open strings between the $N_c$ D4 branes
themselves and an adjoint $N=1$ chiral multiplet $\Phi$, whose two
real scalar components correspond to
fluctuations in the $x^4, x^5$ directions.
Seperating the D4 branes in the $x^4,x^5$
directions correponds to going to a Coulomb branch of the $N=2$ theory.

If we put a set of $N_f$ D6 branes at definite values of $x^4, x^5,
x^6$ there are $N_f$ hypermultiplets of $N=2$ supersymmetry (or $N_f$
pairs of chiral multiplets) in
the fundamental representation of the gauge group $U(N_c)$) coming
from strings between the D4 and D6 branes. Recall that a
hypermultiplet is equivalent to a pair of chiral multiplets $Q$ and
$\tilde{Q}$ in $N=1$ language transforming in the fundamental and the
anti-fundamental representation of the gauge group, respectively.\\

\begin{figure}[ht]
\hbox to\hsize{\hss
\epsfysize=4cm
\epsffile{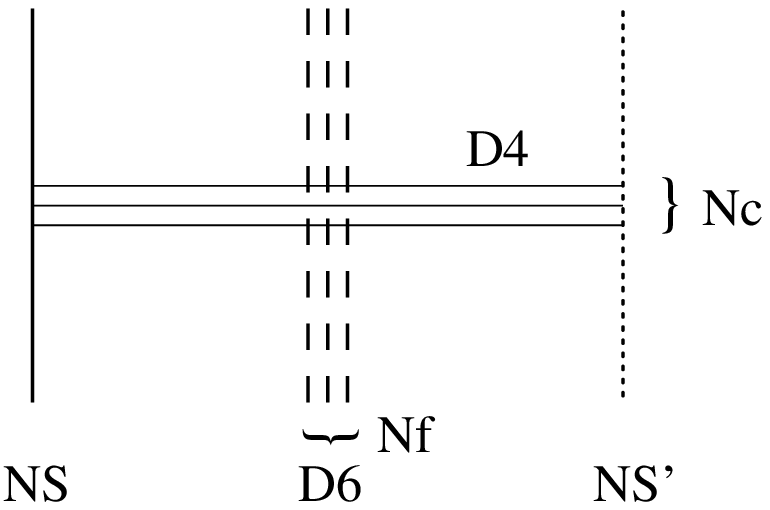} \hss}
\vspace{0cm}
\caption{} \label{pic1}
\end{figure}

\underline{$N=1$ configurations} (cf. figure \ref{pic1})\\

Another possibility is a configuration with one NS and
one NS$^\prime$ at definite values of $x^6, x^7, x^8, x^9$ and $x^4,
x^5, x^6, x^7$, respectively, seperated in the $x^6$ direction. The
difference from the $N=2$ configuration is that the positions of the
$N_c$ D4 branes are now completely fixed.
We cannot seperate the D4 branes and there is no adjoint chiral field.
The world-volume theory on the D4 branes at long distances is a $N=1$
supersymmetric Yang-Mills theory with gauge group $U(N_c)$ \cite{egk}.

If we put a set of $N_f$ D6 branes at definite values of $x^4, x^5,
x^6$ there are $N_f$ pairs of chiral (quark)
multiplets $Q$ and $\tilde{Q}$ in the
(anti)fundamental representation of the gauge group $U(N_c)$ due to
strings between the D4 and D6 branes \cite{egk}. \\
In section 4.3 we discuss generalization  of this configuration to  a model
with a product gauge group.
For both, the $N=2$ and $N=1$ configurations,
moving the D6 branes away from the D4 branes in the $x^4, x^5$ directions
corresponds to mass terms for the quarks in the
world-volume theory which are of the form $m Q \tilde{Q}$.\\

\underline{N=0 configurations}\\

Brane configurations which may be relevant to the study of
field theories without supersymmetry will be the subject of section 5.\\

\underline{Global symmetries}\\

We introduced $N_f$ D6 branes which produced matter multiplets on the D4
brane world-volume theory corresponding to strings between the D4 and
D6 branes. But there are also strings between the D6 branes themselves
which give rise to an $U(N_f)$ gauge theory on the D6
world-volume. Since the D6 branes are much heavier than the D4 branes (they
extend in two additional dimensions) to an D4 brane observer the
$U(N_f)$ gauge theory on the D6 branes appears as a global symmetry -
the $U(N_f)$ flavor symmetry. The $U(1)$ subgroup of the $U(N_f)$
flavor group is the $U(1)_B$ symmetry associated with the conservation
of baryon number. In the corresponding field theory the actual classical
flavor symmetry of the massless $N=1$ theory  is $U(N_f)\times U(N_f)$.
In the $N=2$ case this symmetry is broken explicitly by  the $N=2$
superpotential  to the diagonal $U(N_f)$ symmetry. 

The $N=1$ and the $N=2$ configurations are invariant under rotations
$SO(2)_{45} \sim U(1)_{45}$ in the $x^4, x^5$ plane which corresponds to the
$U(1)_R$ symmetry. The $N=2$ configurations have a rotation symmetry
$SO(3)_{789}$ in the $x^7, x^8, x^9$ directions whose double cover
corresponds to the $SU(2)_R$ symmetry of the $N=2$ supersymmetry
algebra. The $N=1$ configurations are invariant only under its Abelian
subgroup $U(1)_{89}$ which we will denote as the $U(1)_J$ symmetry.
(This $U(1)$ symmetry is also present in certain brane configurations without
supersymmetry which are discussed in section 5).

To summarize
we list the fields and parameters with their
transformation properties\footnote{e.g. the mass parameter $m$
corresponds to moving the D6 in the $x^4, x^5$ directions.
Therefore it transforms as a vector of $SO(2)_{45} \sim U(1)_{45}$ and
we assign charge 2 to it. Spinors of $U(1)_{45}$ have charge 1.}. $M$,
$q$ and $\tilde{q}$ are the meson and the (anti)quark fields,
respectively, that will appear in later sections in
the magnetic dual theories:

\be
\begin{array}{lccc}
             & U(1)_{45} = U(1)_R & U(1)_{89} = U(1)_J & U(1)_B \\[2mm]
W_\a         & 1                  & 1                  & 0      \\
\Phi         & 2                  & 0                  & 0      \\
Q,~\tilde{Q} & 0                  & 1                  & R_Q,~-R_Q  \\
q,~\tilde{q} & 0                  & 1                  & R_q,~-R_q  \\
M            & 2                  & 0                  & 0      \\
m            & 2                  & 0                  & 0      \\
\end{array}
\eel{qunumber}\\[2mm]

Note that the values of the quantum numbers with respect to $U(1)_R$
and $U(1)_J$ are only the classical values. In the quantum theory
the charges are different and only a combination of the two $U(1)$
symmetries remains unbroken. This means that string loop/instanton
effects should break the $U(1)_{45} \times U(1)_{89}$ to a single
unbroken $U(1)_\RR$. The quantum mechanical assignments of the various
charges will be given in a particular case in section 4.3.\\

\underline{Creation and annihilation of D4 branes}\\

Whenever a D6 brane moves in the $x^6$ direction towards an NS brane
and passes through it a new D4 brane stretchinng between the NS and the
D6 brane is
created \cite{egk}. Note that the NS and D6 brane cannot be seperated
in any other direction. In contrast a NS$^\prime$ and a D6 brane
can avoid each other in the $x^4, x^5$ directions.
Such a process was first
studied in \cite{han} in type IIB string theory. A reversed process in
which a D4 brane stretched between a NS and a D6 brane is annihilated;
we will encounter it later.

In \cite{han} the nature of this creation/annihilation process was
made more precise using linking numbers assigned to each brane. The
total linking number of each five-brane that is invariant under any move is the
total magnetic charge measured on that five-brane.
It is defined as the sum of the linking number and the
number of D4 branes with given orientation ending on it. A D4
brane boundary looks like a magnetic charge on the D6 brane. This can
be seen using the results of
\cite{str} where it was shown that in IIB string theory a D
string can end on a D3 brane and that the endpoint of the D string on
the D3 brane is a magnetically charged particle. T-dualizing in
three transverse directions we get a configuration of a D4 brane
ending on a D6 brane in IIA string theory. If a D6 brane
moves through a NS brane its linking number changes by one unit and
consequently a D4 brane is created leaving the total magnetic
charge invariant. The total magnetic charge
measured on a D6 and a NS brane is \cite{han}
\bea
L_{{\rm NS}} & = & \frac{1}{2} (r_{{\rm D6}} - l_{{\rm D6}}) + (L - R) ~,\\
L_{{\rm D6}} & = & \frac{1}{2} (r_{{\rm NS}} - l_{{\rm NS}}) + (L - R) ~,
\eea
where $r$ and $l$ denote the number of branes of the specified type on
the right and on the left of the brane, $L$ and $R$ is the number of
D4 branes ending to the left and to the right. The above expressions
are not modified by the presence of NS$^\prime$ branes.\\[5mm]

\begin{figure}[ht]
\hbox to\hsize{\hss
\epsfysize=4cm
\epsffile{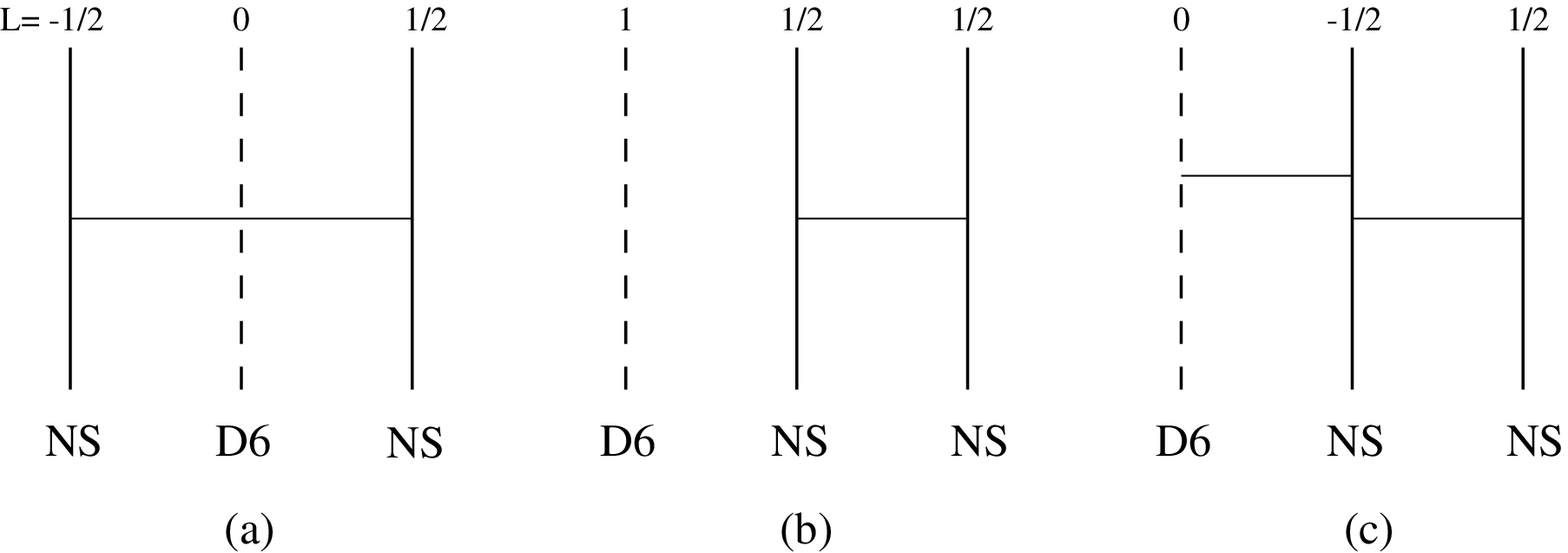}\hss}
\vspace{0cm}
\caption{} \label{pic3}
\end{figure}

In figure \ref{pic3} an example of a creation process is depicted. We
have indicated the total magnetic charges on top of the
branes. The starting configuration figure \ref{pic3}(a) consists of
two NS branes connected by a D4 brane, and a D6 brane between the two
NS branes. This corresponds to an $U(1)$ $N=2$ supersymmetric gauge
theory with one charged hypermultiplet which comes from an open string
between the D4 and the D6 brane. If we move D6 through the left NS
brane we would naivly expect to get the configuration shown in figure
\ref{pic3}(b). This corresponds to a $U(1)$ theory without
hypermultiplet. A careful analysis \`a la \cite{han} shows that the
final configuration must be as in  figure \ref{pic3}(c) where a D4 brane
between the left NS and the D6 brane is created in the
transition. We obtain the same theory on the D4 brane world-volume
although the hypermultiplet now corresponds to an open string between
the left and the right D4 brane.
A geometrical interpretation of this process has recently been given in
\cite{ov}.

\section{$N=2$ Gauge theories and branes}

We start with a configurations of D6, D4 and NS branes.
Recall that in this case only $3/4$ of the
supersymmetry is broken and there is $N=2$ supersymmetry on the D4
brane world-volume. We start with a configuration of two NS
branes and $N_c$ D4 branes stretched between them. The world-volume
theory on the D4 branes is $N=2$ supersymmetric Yang-Mills with gauge
group $U(N_c)$. The D4 branes reside at definite values of $x^7,x^8,x^9$
but their positions in the $x^4, x^5$ directions are allowed to
fluctuate. This fluctuation corresponds to the adjoint (complex) scalar of the
$N=1$ chiral multiplet $\Phi$. If we move the D4
branes in these directions we give vevs to the complex scalar field and we are
in the Coulomb branch of the $N=2$ gauge theory. As discussed in the
previous section this configuration is
symmetric under rotations in the $x^4, x^5$ directions
$U(1)_{45} = U(1)_R$.
In the quantum
field theory this symmetry is broken by instantons to some
discrete subgroup. The $SU(2)_R$ symmetry of the $N=2$ algebra comes
from the rotational invariance in the $x^7, x^8, x^9$ directions.

By putting $N_f$ D6 branes between the two NS branes we can
add $N_f$ hypermultiplets $H_{N=2} = (Q,\tilde{Q})_{N=1}$ in the
fundamental representation of the gauge group.

\begin{figure}[ht]
\hbox to\hsize{\hss
\epsfysize=4cm
\epsffile{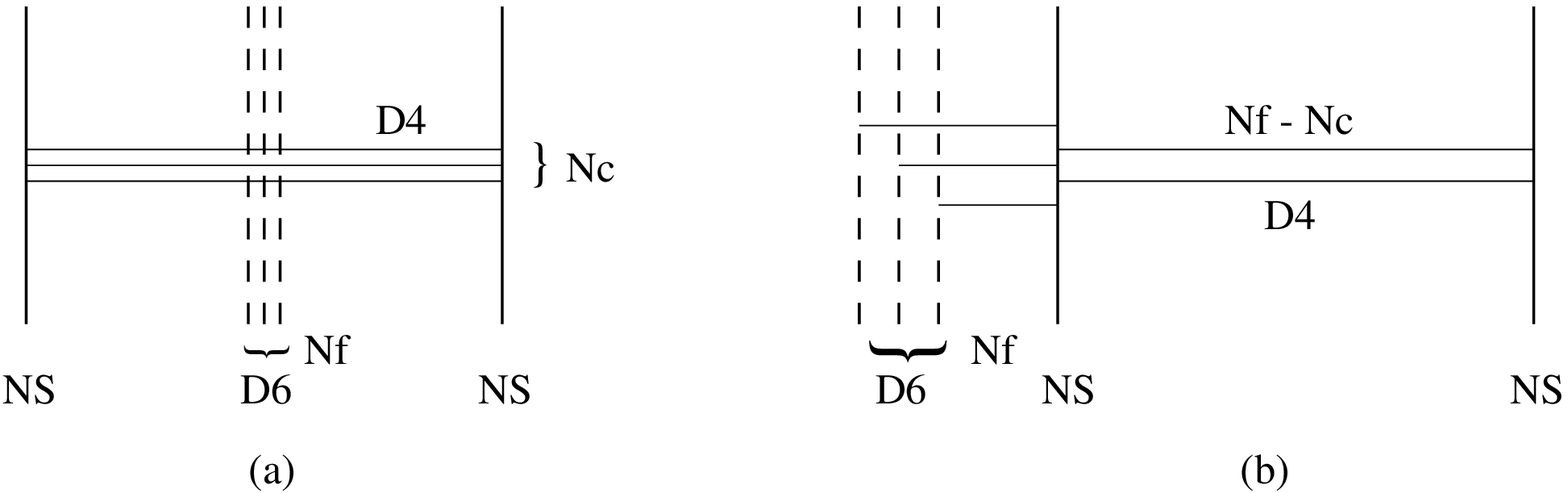}\hss}
\vspace{0cm}
\caption{} \label{pic4}
\end{figure}

Moving the D6 branes in the $x^4, x^5$ directions corresponds to mass
terms for the hypermultiplets $H$. Furthermore if more than one D6
brane touches a D4 brane the D4 brane splits into pieces which can
move between the D6 branes independently in the $x^7, x^8, x^9$
directions. This corresponds to giving a vev to a quark where the vev is
proportional to the separation. The fourth scalar in the $N=2$ hyper
multiplet corresponds to a Wilson line in the $x^6$ direction,
namely the $A_6$ component of the gauge field on the D4 brane.
This reduces, in the field theory picture,  the  
$U(N_c) \times SU(N_f)$ group  to $U(N_c-1) \times SU(N_f-1)$.

We now want to establish the duality of the Higgs branches of the
theory with gauge group $U(N_c)$ and $N_f$ flavors and the theory with
dual gauge group $U(N_f - N_c)$ and $N_f$ flavors \cite{ant}. For that
purpose we perform certain moves in the brane configuration which are
the same as in \cite{egk}.

\begin{enumerate}

\item The starting configuration (figure \ref{pic4}(a)) consists
  of two NS branes separated in the $x^6$ direction, $N_c$ D4 branes
  stretched between them with $x^4 = x^5 = 0$ (i.e. the vev of the
  scalar field in the vectormultiplet is zero) and $N_f$ D6 branes
  which touch the D4 branes $x^4 = x^5 = 0$ (the mass terms are zero
  since we want to study the Higgs phase). We move the D6 branes to the left,
  crossing the left NS brane. In this process we create $N_f$ D4
  branes stretched between the D6 branes and the left NS brane.

\item
  Out of these D4 branes $N_c$
  can connect with the D4 branes stretched between the two NS branes
  (this requires that $N_f \ge N_c$)
  leaving us with $N_c$ D4 branes between the D6 brane and the right
  NS, and $N_f - N_c$ D4 branes stretched between the D6 and the left
  NS brane. The left NS brane can now move in the $x^7, x^8, x^9$
  direction relative to the right NS brane.

\item The left NS brane moves in the $x^6$ direction to the other
  side of the right NS brane without meeting in space-time. Nothing
  special happens in this step.

\item The NS brane comes back to its original position in the
  $x^7, x^8, x^9$ directions, the $N_f - N_c$ D4 branes connecting to
  $N_f - N_c$ D6 branes touch the second NS brane and split into  $N_f
  - N_c$ D4 branes stretched between the two NS branes and  $N_f -
  N_c$ D4 branes stretched between  $N_f - N_c$ D6 branes and the NS
  brane (figure \ref{pic4}(b)).

\end{enumerate}

In the final configuration there are $N_f$ D4 branes stretched between
the $N_f$ D6 branes and a NS brane, and $N_f - N_c$ D4 branes
connecting the two NS branes. The world-volume theory on the D4 branes
is a $N=2$ $U(N_f - N_c)$ gauge theory with $N_f$ hypermultiplets from
open strings connecting the two sets of D4 branes. Since the
positions of the $N_f$ D4 branes between the D6 branes and the NS
brane are completely fixed there are no massless states coming from
open strings between these D4 branes (no mesons), as expected from
field theory.

\begin{figure}[ht]
\hbox to\hsize{\hss
\epsfysize=4.5cm
\epsffile{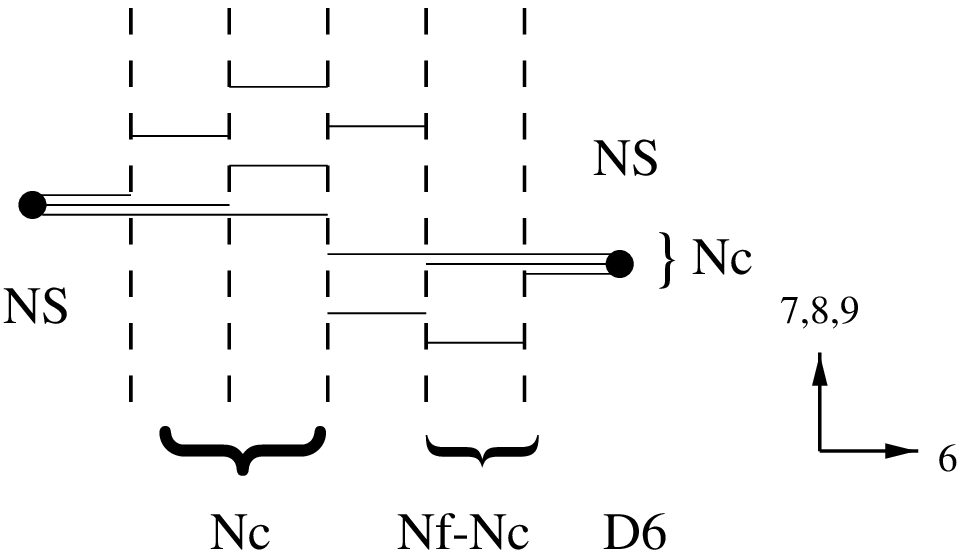}\hss}
\vspace{0cm}
\caption{} \label{pic4a}
\end{figure}

There is an alternative way to see that the NS branes can move in the
$x^7, x^8, x^9$ directions which corresponds to a FI term in the field
theory. Remember that this feature is crucial to make all the moves to
obtain the dual brane configuration.
The main point is to split the D4 branes as
much as possible already in the starting position i.e. we choose a
generic point in the Higgs moduli space. In order to do so the D6
branes must touch the D4 branes which allows the D4 branes to
split and the parts between two D6 branes can move in the $x^7, x^8,
x^9$ directions. Note that there is sublety for D4 branes stretched between an
NS and an D6 brane which is explained in the next paragraph.
Only if $N_f \ge N_c$ can all the D4 branes split and
the two NS branes are free to move with respect to each other
(together with the D4 branes
attached to them) in the $x^7, x^8, x^9$ directions (figure
\ref{pic4a}). The relative offset between the two NS branes
corresponds to turning on FI terms in the field theory on the brane.
On the Higgs branch FI terms do not  break supersymmetry.

Let us end this section with a comment on the dimension of the Higgs
branch. To count the number of parameters we split the D4 branes
between the D6 branes as much as possible. Remember that the D4 brane
stretched between an NS and a D6 brane cannot move whereas a D4
brane between two D6 branes can move. Each D4
brane which is allowed to move gives two complex parameters (the
$x^7, x^8, x^9$ directions and a Wilson line in the $x^6$ direction). The naive
counting would then give $2 N_c (N_f - 1)$ which contradicts our
expectation from field theory where the dimension is $2 N_c (N_f -
N_c)$. To remedy this situation recall the discussion of  the  so called
$s$-configurations in \cite{han}. The main point is that if we have a NS
and a D6 brane we get a supersymmetric vacuum only if there is just
one D4 brane stretched between them.
So we have to be more careful in the splitting of the D4
branes and we have to avoid such $s$-configurations. The resulting
picture is given in figure \ref{pic4a} and the counting now gives the
correct dimension of the Higgs branch $2 N_c (N_f - N_c)$. This is invariant
under $N_c \leftrightarrow N_f - N_c$.

\section{Dualities in $N=1$ theories}

\subsection{Seiberg's Duality}

The relevant configuration was proposed in \cite{egk} and is drawn
in figure \ref{pic1}. The NS brane is located at $x^6 = x^6_{{\rm NS}} <
x^6_{NS^\prime}$ and $x^7 = x^8 = x^9 = 0$, the NS$^\prime$ brane is at
$x^6 = x^6_{{\rm NS}^\prime} > x^6_{{\rm NS}}$
and $x^4 = x^5 = x^7 = 0$. The D4
branes reside at $x^4=x^5=x^7=x^8=x^9=0$ and are extended in the $x^6$
direction with $x^6_{{\rm NS}} < x^6 < x^6_{{\rm NS}^\prime}$.
For the moment the D6
branes are chosen to reside at $x^4=x^5 = 0$. We will discuss deformations of
this configuration later. The $x^6$ positions of the D6 branes are somewhere
between the NS and the NS$^\prime$ brane. This means that the D6
branes touch the D4 branes and the quarks which correspond to open
strings between the D4 and the D6 branes are massless. As was shown in
\cite{egk}, to get to the
magnetic description we have to go through the following four moves.

\begin{itemize}

\item We move all D6 branes to the left of the NS brane. When the
  $x^6$ positions of a D6 and a NS brane coincide they actually meet
  in the $x^0, x^1, x^2, x^3$ directions. By passing through the NS
  brane a D4 brane is generated for each D6 and this D4 brane is
  stretched between the NS and the D6 brane. The world-volume theory on
  the D4 branes does not change, only the (anti)quarks correspond now
  to open strings between the two sets of D4 branes.
  Now $N_c$ D4 branes of the $N_f$
  stetched between the D6 branes and the NS brane can be connected
  with the $N_c$ D4 branes stretched between the NS and the
  NS$^\prime$ brane. For this it is important that $N_f \ge N_c$,
  otherwise there is always at least one D4 brane between NS and
  NS$^\prime$ which would force the $x^7$ positions of the NS and the
  NS$^\prime$ to be equal. Thus we have $N_c$ D4 branes between the D6 and the
  NS$^\prime$, and $N_f - N_c$ D4 branes between the D6 and the NS.

\item  Since none of the D4 branes ending on NS$^\prime$ also ends
  on the NS, the NS brane can be lifted in the $x^7$ direction away from the
  NS$^\prime$. This corresponds to turning on a Fayet-Ilioupulos
  D-term for the $U(1)$ factor of the gauge group. Since we have
  turned off the massterms we are in the Higgs branch and in this case
  the FI D-term does not break supersymmetry.

\item Now the NS brane can move to the other side of the NS$^\prime$
  brane such that they avoid each other. Nothing special happens in
  this step.

\item Next we turn off the FI D-term which means that the $x^7$ position of
  the NS brane returns to its original value. This implies that the
  NS$^\prime$ touches the $N_f - N_c$ D4 branes and they can
  split. The final configuration is therefore $N_f$ D4 brane between
  the D6 and NS$^\prime$ brane and $N_f - N_c$ D4 branes between NS
  and NS$^\prime$ (figure \ref{pic2}). The matter content is equal to
  Seiberg's magnetic theory. The gauge group is $U(N_f - N_c)$, there
  are $N_f$ magnetic  (anti)quarks) in the (anti)fundamental 
   of the
  gauge group coming from strings between the two sets of D4 branes, and
  the magnetic mesons which arise from strings between the $N_f$ D4
  branes between the D6 branes and the NS$^\prime$.
\footnote{Near the origin of the moduli space the correct
  identification of the mesons involves also mixing with states
  arising from strings between the $N_f$ D4 branes and the D6
  branes. We will discuss this point later. We would like to thank
  O. Aharony for discussions on this issue.}

\begin{figure}[ht]
\hbox to\hsize{\hss
\epsfysize=4cm
\epsffile{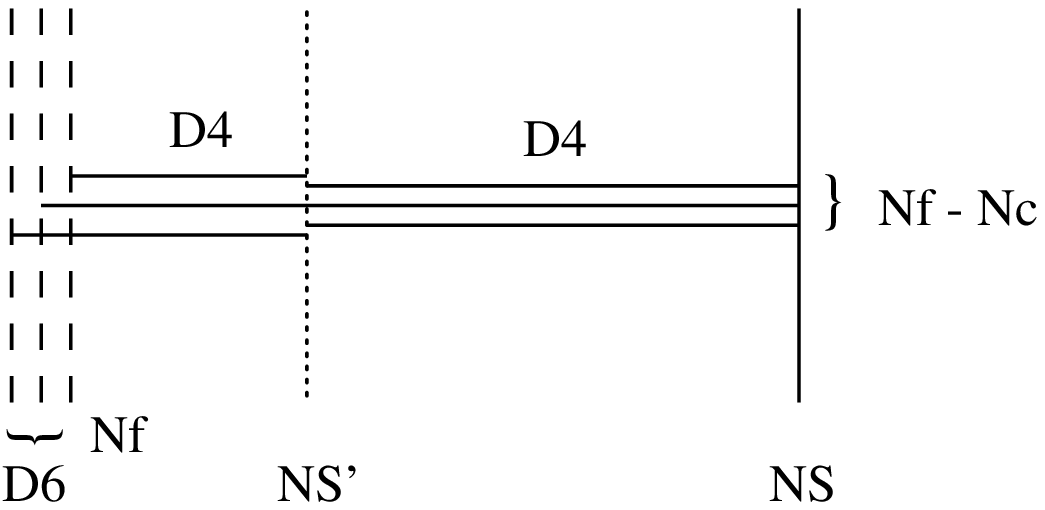}\hss}
\vspace{0cm}
\caption{} \label{pic2}
\end{figure}

\end{itemize}

\begin{figure}[ht]
\hbox to\hsize{\hss
\epsfysize=4cm
\epsffile{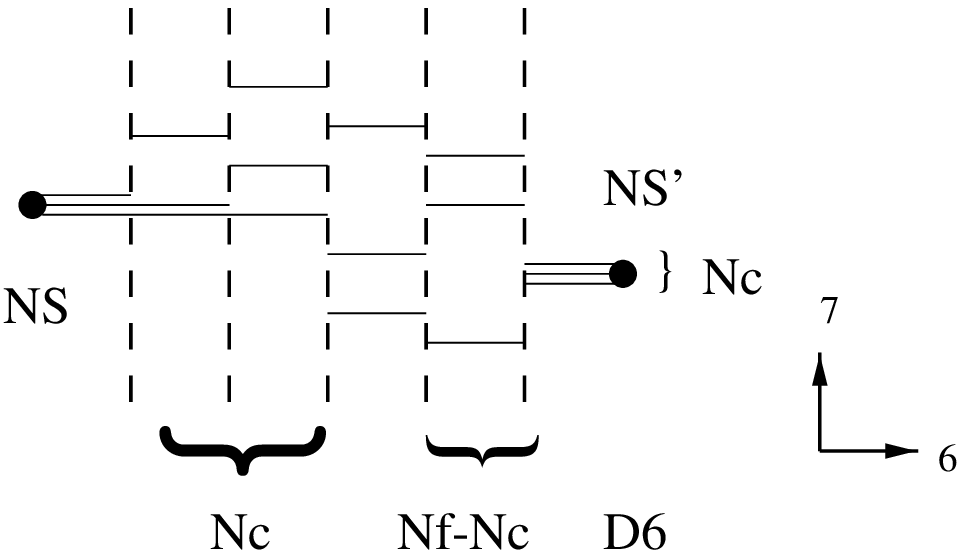}\hss}
\vspace{0cm}
\caption{} \label{pic4b}
\end{figure}

To gain more insight into the issue of the FI term it is useful to
start at a generic point of the Higgs branch.
This means that in the starting position the D6
branes touch the D4 branes which split into two parts. The position of
the left part of a D4 brane is always fixed, but the right part can
move in the $x^8, x^9$ directions. To avoid $s$-configurations only
one D4 brane part is allowed to stretch between the NS and one specific D6
brane, but between two D6 branes there can be more than one D4
branes separated in the $x^7, x^8, x^9$ directions.
There is also the possibility of D4 brane parts between D6 branes and
the NS$^\prime$ brane which can move in the $x^8, x^9$ directions.
Complete splitting is
only possible for $N_f \ge N_c$ and once this is achieved the NS brane can
move together with the D4 branes attached to it in the $x^7, x^8, x^9$
directions (fig. \ref{pic4b}). Out of the three parameters only the separation
 of the NS and
the NS$^\prime$ in the $x^7$ direction corresponds to the FI term since the
world-volume of the NS$^\prime$ brane extends in the $x^8, x^9$ directions.

We now want to study in turn two different deformations of the
theories in the electric and the magnetic description. We recall that the
correspondence between the deformed electric and magnetic theories
formed the basis for the belief in the duality within the field theory
context.\\

\underline{Mass terms:}\\

\begin{figure}[ht]
\hbox to\hsize{\hss
\epsfysize=4cm
\epsffile{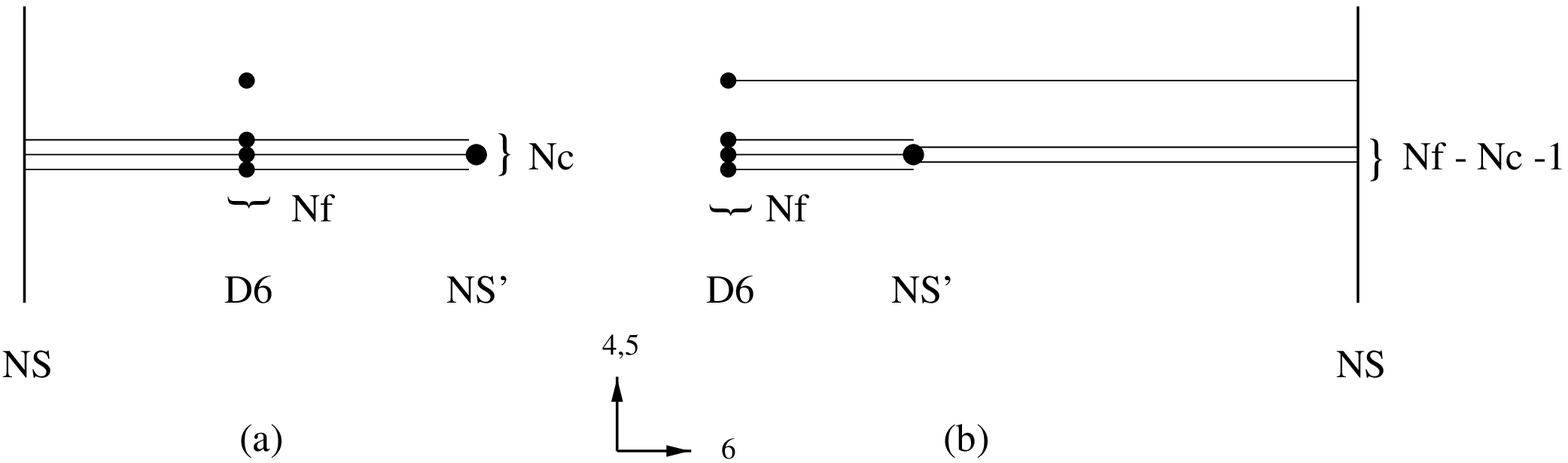}\hss}
\vspace{0cm}
\caption{} \label{pic5}
\end{figure}

Giving a mass to one of the electric
quarks reduces the flavor symmetry $U(N_f)$ to $U(N_f - 1)$.
The field theories on the D4 branes have a  classical flavor group bigger than
$U(N_f)$, namely
$U(N_f) \times U(N_f)$ but in the brane context we see only the
diagonal part of it which arises from the D6 brane sector.
In the magnetic theory giving a mass to one of the mesons induces a non-zero
vev for one of the magnetic quarks via the equations of motion of the
massive meson. Thus the magnetic gauge group is reduced to $U(N_f - N_c - 1)$. 
What does this correspond to in the brane configuration?
The starting point is a configuration of one NS and one NS$^\prime$
brane, $N_c$ D4 branes stretched between NS and NS$^\prime$, all on
top of each other and $N_f$ D6 branes. Moving one of the D6 branes in
the $x^4, x^5$ directions away from the D4 branes (figure
\ref{pic5}(a)) gives mass to a quark in the world-volume theory and
reduces the global symmetry to $U(N_f - 1)$. The gauge symmetry in the
electric theory remains unbroken.\\[5mm]

The final configuration is the following. $N_f - N_c - 1$ D4 branes
stretched between NS and NS$^\prime$, $N_f - 1$ D4 branes between the
D6 branes and the NS$^\prime$ brane, and a D4 brane between
one D6 and the NS brane separated from the other D4 branes in the
$x^4, x^5$ directions  (figure \ref{pic5}(b)). This corresponds to a
mass term for one of the mesons (say $M^{N_f}_{\tilde{N_f}}$) which,
via the equations of motion, gives a vev to a magnetic quark. This
reduces the gauge group and the global flavor symmetry. To obtain this
configuration from the original configuration of the magnetic theory
we have to connect a D4 brane between D6 and NS$^\prime$ with a D4
brane between NS and NS$^\prime$. Only after this connection the
resulting D4 brane can be lifted off from NS$^\prime$ in the $x^4,
x^5$ directions. \\

\underline{Vevs for quarks:}\\

\begin{figure}[ht]
\hbox to\hsize{\hss
\epsfysize=4cm
\epsffile{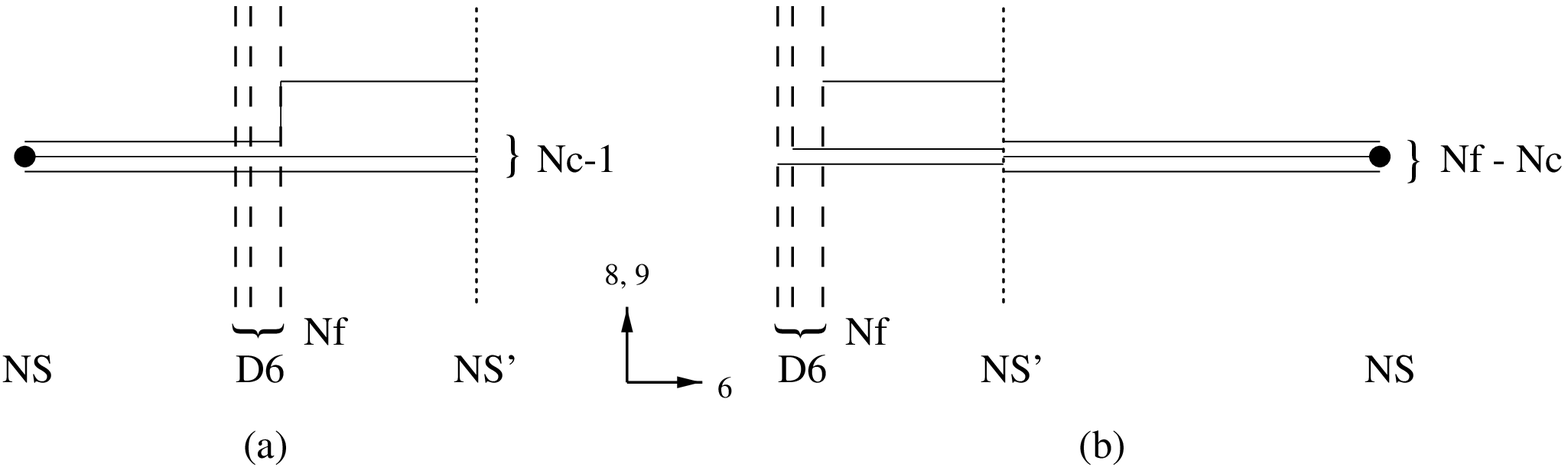}\hss}
\vspace{0cm}
\caption{} \label{pic6}
\end{figure}

The initial configuration is again one NS brane on the left side and one
NS$^\prime$ brane on the right, $N_c$ D4 branes stretched between NS
and NS$^\prime$, all coincident and $N_f$ D6 branes with
$x^6$ positions somewhere between the NS and the NS$^\prime$ brane. If
we tune the $x^4, x^5$ postions of one of the D6 branes such that it
touches the D4 branes one (or more) of the D4 branes can split and the
part between the D6 and the  NS$^\prime$ brane can move in the $x^8,
x^9$ directions (figure \ref{pic6}(a)). We assume that one D4 brane
splits on one specific D6 brane. If more than one D4 brane split on a
specific D6 brane there would be more than one D4 branes between NS and
this specific D6. This corresponds to an $s$-configuration \cite{han}
and breaks supersymmetry which we want to avoid.
The separation in $x^8, x^9$ corresponds to giving a vev to
one of the quarks. Thus
the gauge group in the electric theory is reduced to $U(N_c - 1)$ and
the number of massless (anti)-quarks is $N_f - 1$.\\[5mm]

The remaining $N_c - 1$ D4 branes do not split and going through the usual
steps we obtain the final configuration figure \ref{pic6}(b). There
are $N_f - N_c$ D4 branes between NS and NS$^\prime$ and $N_f - 1$ D4
branes between the D6 branes and the  NS$^\prime$ brane. The
world-volume theory on the D4 branes is $N=1$ supersymmetric gauge
theory with gauge group $U(N_f - N_c)$, $N_f-1$ magnetic quarks and
mesons in the $(N_f-1,\overline{N_f-1})$ representation. The fate of the
split D4 brane deserves some comments. The piece that moved in the $x^8, x^9$
directions is stretched between one of the $N_f$ D6 branes and the
NS$^\prime$ brane and it corresponds to a meson which has been given a
vev proportional to the distance to the other $N_f-1$ D4 branes in the
$x^8, x^9$ directions (figure \ref{pic6}(b)). The other part which,
in the initial position, is stretched between NS and one of the D6
disappears as the D6 brane moves through the NS brane and no additional
D4 brane is created. In this ``annihilation"
process the number of D4 branes has to be
reduced  by one per D6 brane. This
annihilation process is just the reverse of the creation process of D4
branes, illustrated in figure 3.

We recall that giving a vev to the mesons still should leave us with
$N_f^2$ massless mesons, some of them correspond to the Goldstone
bosons associated with the breaking of the flavor symmetry. 
 Moving one of the D4 branes in the magnetic
picture along the $x^8, x^9$ directions corresponds to giving a vev to
say $M_{1}^{1}$. Naively, it would seem as if we have now only $(N_f -
1)^2$ massless mesons, since the strings between the D4 brane which we
have moved and the rest of the $N_f - 1$ D4 branes are all long. The
resolution of this apparent paradox is that the mesons are really a
mixture of D4-D4 and D4-D6 string states. (We recall, however, that
there are no strings between D6 and D4 which terminate on each
other). The coefficients of this mixture depend on the particular
point in the moduli space. At a generic point where all D4 branes and
D6 branes are well separated the mesons are essentially D4-D4
strings. However, near the origin of the moduli space there must be a
substential component of the states coming from strings between
D4 and D6 branes which do not terminate on each other.

We found that moving D6 branes or splitting D4 branes has an
interpretation in field theory as mass terms or vevs for
fields. Furthermore we were able to find a direct map between the
deformations in the electric and magnetic theory which provides
further evidence that the brane configuration figure \ref{pic1}
reproduces Seiberg's duality.
In the brane picture it is straightforward to see that the dual of the
dual theory is the original one. All the steps have to be done
in reversed order and we end up in the initial configuration. Recall
that in field theory it is much more involved to show it \cite{sei1}.

\subsection{Adjoint Duality}

In \cite{kut} dualities between theories with one chiral superfield in
the adjoint representation was proposed. The electric theory contains
an $U(N_c)$ vector multiplet, $N_f$ (anti-) quarks $Q_i$ and
$\tilde{Q}^{\tilde{i}}$ in the (anti) fundamental representation of
the gauge group and an adjoint $X$. Adding the following
superpotential breaks $N=2$ to $N=1$ supersymmetry:
\be
W_e = {\rm Tr} \sum_{l=1}^{k+1} c_l X^l ~ , ~ c_{k+1} = 1.
\ee
The dual theory is an $U(k N_f - N_c)$ gauge theory with magnetic
quarks $q^i$ and $\tilde{q}_{\tilde{i}}$, an adjoint field $Y$ and a
set of $k$ mesons $M_l ~, ~ l=1,\ldots,k$. These mesons correspond to
the composite operators $ Q X^{l-1} \tilde{Q} ~,~ l = 1,\ldots,k$ in the
electric theory. The magnetic superpotential has the form:
\be
W_m = {\rm Tr} \sum_{l=1}^{k+1} d_l Y^l + \sum_{l=1}^k M_l \tilde{q}
Y^{k-l}q ~ , ~ d_{k+1} = 1.
\ee
The exact map between the $c_l$ and $d_l$ has been worked out in
the third reference of \cite{kut}. The electric superpotential has $k$
minima at $\partial W_k/\partial X  = 0$. This means that $X$ has
$k$ different eigenvalues of multiplicity $r_i$ with $\sum_{i=1}^k r_i
= N_c$ which breaks the gauge symmetry to $\prod_{i=1}^k U(r_i)$. In
the following we assume that the adjoint is massive.
We also note that there is never an
adjoint flat direction due to the fact
that the gauge group has an additional $U(1)$ factor.
(For  the $SU(N)$ gauge symmetry case a flat direction appears for even $N$
see e.g. \cite{asy})

A brane configuration representing this gauge theory was proposed in
\cite{egk} and is the following. Instead of one there are $k$ NS
branes whose $x^8, x^9$ positions correspond to the deformation
parameters $c_i$ of the electric superpotential. Furthermore there is
one NS$^\prime$ brane and the $i$-th NS brane is connected to the NS
brane by $r_i$ D4 branes, and there are $N_f$ D6 branes which are
needed to generate the matter (quark) content. Notice that there is no
direct way to {\it see} the adjoint field $X$ directly as open
strings, but there is a direct relation between the NS positions which
are equal to the eigenvalues of the adjoint chiral superfield $X$ and
the deformation parameters $c_l$ of the superpotential.
The fact that we do not see $X$ as a fluctuation in the brane configuration
is correlated with the observation that for a $U(N)$ gauge theory there 
is never
an adjoint flat direction (which, if existed, would have had to correspond 
to some brane motion). 

\begin{figure}[ht!]
\hbox to\hsize{\hss
\epsfysize=5cm
\epsffile{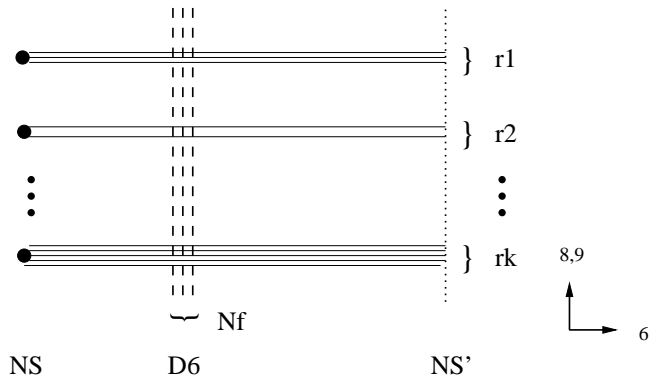}\hss}
\vspace{0cm}
\caption{A configuration with $k=3$} \label{pic7}
\end{figure}

This configuration consists of $k$ copies of the configuration studied
in the previous section and it is easy to guess what the final
configuration corresponding to the magnetic theory must be. There are
$k$ sets of $N_f$ D4 branes stretched between the D6 branes and the
NS$^\prime$, and $k$ sets of $r_i$ D4 branes connecting the $i$-th NS
with the NS$^\prime$ brane. The resulting gauge group of the
world-volume theory is broken: $U(k N_f - N_c) \to \prod_{i=1}^k U(N_f
- r_i)$ and there are $N_f$ magnetic (anti)-quarks in the fundamental
of the broken gauge group. The $k$ mesons come from the $k$ sets of D4
branes between the D6 branes and the NS$^\prime$.
Again, near the origin of the moduli space appropriate
mixing with D4-D6 string states occures as discussed in the case of
Seiberg's duality.

It presents no difficulty to realize the possible deformations in
these theories: mass terms and vevs. It is completely analogous to the
{\rm moves} in the section on Seiberg's duality. \\

\underline{Mass terms}\\

A mass term for one quark in the electric theory corresponds to lifting
one of the $N_f$ D6 branes in the $x^4, x^5$ directions making the
strings between the D4 branes and this D6 brane long and thus
massive. The flavor symmetry is reduced to $SU(N_f - 1)$ and the electric
gauge group is unchanged.

The final configuration is the following. $k$ groups of $N_f - 1$ D4
branes connecting the D6 branes to  NS$^\prime$ which are separated in
the $x^8, x^9$ directions. There are $k$ groups of $N_f - r_i - 1$ D4
branes connecting NS$^\prime$ and the NS$_i$ branes. The D6 brane
which is lifted in the $x^4, x^5$ directions creates $k$ D4 branes
stretched between this D6 and the NS$_i$ branes. In the magnetic
theory this corresponds to giving mass to the $(N_f,\overline{N_f})$
component of
the $k$ mesons. The equations of motion for the massive fields give a
vev to one of the magnetic quarks. The magnetic gauge group is
therefore $U(k N_f - N_c) \to \prod_{i=1}^k U(N_f - 1 - r_i)$ and the
flavor group is broken to $SU(N_f - 1)$.\\

\underline{Vevs for quarks}\\

We consider the case where one of the $r_1$ D4 branes connecting
NS$_1$ and NS$^\prime$ splits into two parts separated in the
$x^8,x^9$ directions. The gauge group is broken,
\be
U(r_1) \otimes U(r_2) \otimes \ldots \to U(r_1 - 1) \otimes U(r_2)
\otimes \ldots ~,
\ee
and the flavor symmetry is reduced to $SU(N_f - 1)$.

In the magnetic theory this corresponds to giving a vev to one
component of one of the mesons. This also gives mass to one of the quarks
due to the tree level superpotential and reduces the flavor group:
$SU(N_f) \to SU(N_f-1)$.
The magnetic gauge group remains unaffected.
We would like to stress again that although we do not see the adjoint
field, it is reassuring that the deformations associated with mesons
containing the adjoint field are accounted for. This gives us
confidence that the brane configuration suggested in \cite{egk} is
indeed producing the ``adjoint duality" \cite{kut}.

\subsection{Duality of theories with product gauge groups}

\begin{figure}[ht!]
\hbox to\hsize{\hss
\epsfysize=10cm
\epsffile{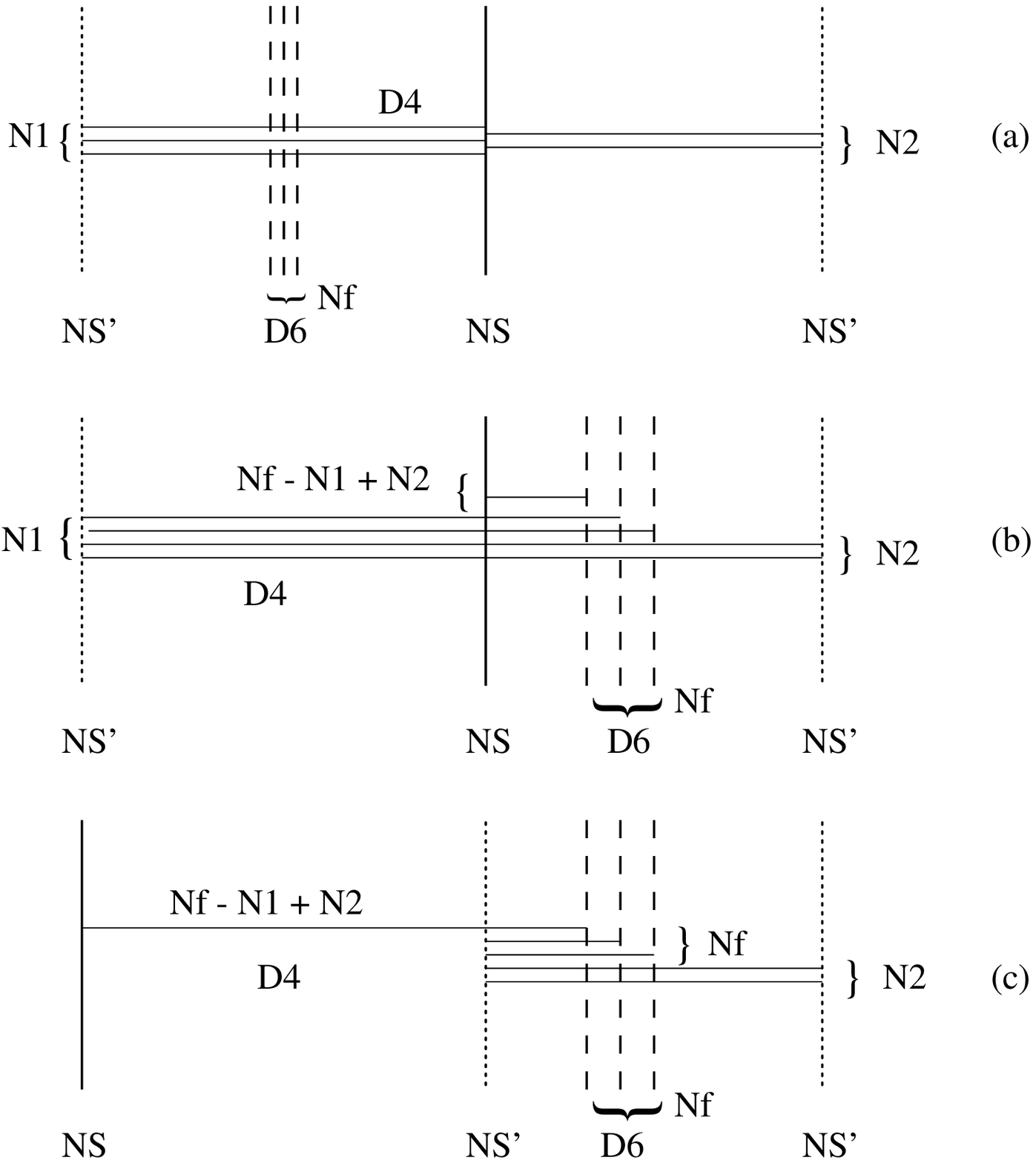}\hss}
\vspace{0cm}
\caption{} \label{pic9}
\end{figure}

In this section we want to study an extension of the brane
configurations studied above. The brane configuration is drawn in
figure \ref{pic9}(a) and consists of one NS brane, two NS$^\prime$ branes
and $N_f$ D6 branes between the NS and the first NS$^\prime$ brane.
In addition there are $N_1$ D4 branes stretched between the NS and the first
NS$^\prime$ brane, and $N_2$ D4 branes between the NS and the second
NS$^\prime$ brane.
We take $N_1\geq N_2$ and
chose the  $x^4,x^5$ values of the two NS$^\prime$ branes to coincide.
The reason for these choices will become clear below.
The world-volume theory on the D4 branes is a $N=1$
supersymmetry gauge theory with gauge group $U(N_1) \times U(N_2)$
with gauge couplings
$1/g_1^2 \sim | x^6_{{\rm NS}^\prime_1} - x^6_{{\rm NS}} |$ and
$1/g_2^2 \sim | x^6_{{\rm NS}}
-x^6_{{\rm NS}^\prime_2} | $. Since there are two
independent gauge fields there are, at least classically, two
$U(1)_\RR$ symmetries. There are $N_f$ (anti-) quarks in the
(anti-)fundamental, $Q$ and $\tilde{Q}$, and chiral multiplets, $N$
and $\tilde N$, in the $(N_1, \overline{N_2})$ and the
$(\overline{N_1}, N_2)$ representations of the gauge group.
They arise from open strings between the two groups of D4 branes.
The global symmetry of the world-volume theory is (recall that only
the diagonal $SU(N_f)$ factor is manifest in the brane picture)

\be
SU(N_f) \times SU(N_f) \times U(1)_J \times U(1)_B \times
U(1)_\RR^{(1)} \times U(1)_\RR^{(2)}~~,
\ee

Of the  $U(1)_J$, $U(1)_\RR^{(1)}$ and $U(1)_\RR^{(2)}$ symmetry only
two combinations remain unbroken in the quantum theory and we will
denote them by $U(1)_\RR^{(1)}$ and $U(1)_\RR^{(2)}$. The quantum
numbers are determined  by the requirement of
gauge anomaly cancellation
and are summarized in the following table:

\be
\begin{array}{lccccccc}
    &U(N_1)&U(N_2) & SU(N_f) & SU(N_f) &
         U(1)_B &U(1)_\RR^{(1)}&U(1)_\RR^{(2)}\\[3mm]
 W_\alpha^{(1)} & (N_1)^2 & 1       & 1 & 1 & 0 & 1 & 1 \\[1mm]
 W_\alpha^{(2)} & 1       & (N_2)^2 & 1 & 1 & 0 & 1 & 0 \\[1mm]
 N         & N_1 & \overline{N_2} & 1 & 1 & 1 \over N_1 & {N_1 - N_2
         \over N_1} & 1 \\[1mm]
 \tilde{N} & \overline{N_1} & N_2 & 1 & 1 & -{1 \over N_1} & {N_1 - N_2
         \over N_1} & 1 \\[1mm]
 Q & N_1 & 1 & N_f & 1 & 1 \over N_1 & N_2^2 - N_1^2 + N_1 N_f \over N_1 N_f
         & N_f - N_1 \over N_f \\[1mm]
 \tilde{Q} & \overline{N_1} & 1 & 1 & \overline{N_f} & -{1 \over N_1} & N_2^2
         - N_1^2 + N_1 N_f \over N_1 N_f & N_f - N_1 \over N_f
\end{array}
\ee\\[5mm]
To find the dual of this theory we follow the same steps as in the case
of Seibergs Duality. We begin by moving all D6 branes to the right of
the NS brane and create, in the by now familiar manner, $N_f$ D4 branes
stretched between the D6 branes and the NS brane. In the creation
process each of these $N_f$ D4 branes inherites its $x^4, x^5$
postions from the D6 brane it ends on and its $x^7, x^8, x^9$
positions from the NS brane. We now connect the $N_2$ D4 branes on the
right to $N_2$ out of the $N_1$ D4 branes on the left.
By adjusting the $x^4, x^5$ positions of $N_1-N_2$ D6 branes,
$N_1-N_2$ of the $N_f$ newly created D4 branes can be connected to
those $N_1-N_2$ D4 branes stretched between the left NS$^\prime$ and
NS brane, which had not yet been connected to the D4 branes on the
right of the NS$^\prime$ brane. (figure \ref{pic9}(b)). Note that this
requires, in addition to $N_1\geq N_2$ also $N_f\geq N_1-N_2$.

Now the NS brane can be lifted in the $x^7$ direction which corresponds
to turning on a FI term in the world-volume field theory. The NS brane
can be moved to the left of the left NS$^\prime$ brane. At this stage
nothing special happens and we can put the $x^7$ position of the NS
brane back to its original value. The left NS$^\prime$ touches the D4
branes stretched between the NS and the other branes and they
split. The final configuration is drawn in figure \ref{pic9}(c).

Let us summarize the field content of the D4 world-volume theory in the
final brane configuration and discuss whether this is a possible $N=1$
duality by comparing the chiral rings of the two theories and checking
the 't Hooft anomaly matching conditions. The theory has $U(N_f -
N_1 + N_2) \times U(N_2)$ gauge symmetry. Furthermore there are two
kinds of quarks, magnetic (anti-) quarks $q$ and $\tilde{q}$ in the
(anti-)fundamental of $U(N_f -N_1 + N_2)$ which come from
open strings between the two sets of D4 branes, and
(anti-) quarks $r$ and $\tilde{r}$ in the (anti-) fundamental of
$U(N_2)$ from strings between D4 and D6 branes. There
are also $N_f^2$ meson fields $M$, from open strings between the $N_f$
D4 branes terminating on the D6 branes,
a chiral superfield $X$ in the adjoint
of $U(N_2)$, whose scalar components are related to the fluctuations of the
$N_2$ D4 branes between the two NS$^\prime$
branes in the $x^8,x^9$ directions,
and, finally, chiral superfields, $n$ and $\tilde n$
in mixed representations of the
gauge group from strings between the left and the right set of D4 branes,
both extending between one NS and one NS$^\prime$ brane.
The classical superpotential of the magnetic theory is
\be
W_m = q M \tilde{q} + n X \tilde{n} + n r \tilde{q} + \tilde{n}
\tilde{r} q ~.
\eel{superpot}
The global symmetry of this theory is the same as for the
original, electric theory. We summarize the quantum numbers under the
local and global symmetries in the following table:
\be
\begin{array}{lccccccc}
 &U(\tilde{N_1})&U(N_2)&SU(N_f)&SU(N_f)&U(1)_B&
                    U(1)_\RR^{(1)}&U(1)_\RR^{(2)}\\[3mm]

 W_\alpha^{(1)} & (\tilde{N_1})^2 & 1       & 1 & 1 & 0 & 1 & 1 \\[1mm]

 W_\alpha^{(2)} & 1               & (N_2)^2 & 1 & 1 & 0 & 1 & 0 \\[1mm]

 n         & \tilde{N_1} &\overline{N_2}& 1    & 1 &
 1 \over \tilde{N_1} & {N_2 \over N_1} & 0 \\[1mm]

 \tilde{n} & \overline{\tilde{N_1}}&N_2 & 1    & 1 &
 -{1 \over \tilde{N_1}} & {N_2 \over N_1} & 0 \\[1mm]

 q & \tilde{N_1}  & 1 & \overline{N_f}& 1 & 1 \over \tilde{N_1} &
      {N_1^2 - N_2^2 \over N_1 N_f}  & {N_1 \over N_f} \\[1mm]

 \tilde{q} & \overline{\tilde{N_1}} & 1 & 1 & N_f & -{1 \over
   \tilde{N_1}} &  {N_1^2 - N_2^2 \over N_1 N_f}  & {N_1 \over
   N_f}  \\[1mm]

 r         & 1 & N_2& N_f    & 1 & 0 & {N_2^2 - N_f N_2 - N_1^2 +
       2 N_f N_1 \over N_f N_1} & {2 N_f - N_1 \over N_f} \\[1mm]

 \tilde{r} & 1 & \overline{N_2}& 1 & \overline{N_f}& 0 & {N_2^2 - N_f
   N_2 - N_1^2 + 2 N_f N_1 \over N_f N_1} & {2 N_f - N_1 \over N_f} \\[1mm]

 M         & 1            & 1          & N_f& \overline{N_f} & 0 &
2 {N_2^2 - N_1^2 + N_1 N_f\over N_f N_1} & 2{N_f -  N_1 \over N_f} \\[1mm]

 X         & 1 &           (N_2)^2     & 1  & 1 & 0 &
2 {N_1 - N_2 \over N_1} & 2
\end{array}
\eel{table2}
with
\be \tilde{N_1} = N_f - N_1 + N_2 ~~.
\ee
With these charge assignments all 't Hooft anomaly matching conditions
for the global symmetries are satisfied.

To provide further evidence for the duality let us compare the
gauge-invariant operators in the two theories. The mesons of the
electric theory may all be identified with gauge singlets of the
magnetic theory (in fact one has to consider $SU(N)$ gauge groups
rather than $U(N)$):

\be
M \sim Q \tilde{Q}~, r \sim Q \tilde{N}~, \tilde{r} \sim \tilde{Q} N~,
X \sim N \tilde{N}~.
\eel{mesons}

The relevant gauge invariant baryonic operators \footnote{e.g. by $(Q)^{N_1}$
we mean $Q^{i_1} \ldots Q^{i_{N_1}}$ where contraction of the color
indices with an totally antisymmetric rank $N_1$ tensor is understood.}
of the electric theory are
$(Q)^{N_1}$, $(\tilde Q)^{N_1}$, $(N)^{N_2} (Q)^{N_1 - N_2}$ and
$(\tilde N)^{N_2} (\tilde Q)^{N_1 - N_2}$, and the corresponding
operators of the magnetic theory are $(n)^{N_2} (q)^{\tilde{N_1} - N_2}$,
$(\tilde n)^{N_2} (\tilde q)^{\tilde{N_1} - N_2}$, $(q)^{\tilde{N_1}}$
and $(\tilde q)^{\tilde{N_1}}$, respectively.
It is easy to see that the operators in the electric and their
corresponding operators in the magnetic theory carry the same $U(1)$
charges and are in the same antisymmetric representation of the flavor
group $SU(N_f)$ with dimensions ${N_f \choose N_1} =
{N_f \choose N_f - N_1}$ and $ {N_f \choose N_1 - N_2} =
{N_f \choose N_f - N_1 + N_2 = \tilde{N_1}} $ for the first two and
the last two operators, respectively.

There is also a straightforward
field theoretic interpretation of this duality. The
starting point is
$N=1$ supersymmetric QCD with gauge group $U(N_1)$ and $\tilde{N}_f
=  N_f + N_2$ (anti-)quarks. Now we can apply Seiberg's
dualtity to this theory and we will find a magnetic theory with gauge
group $U(\tilde{N}_f - N_1) = U(N_f + N_2 - N_1)$, $\tilde{N}_f = N_f
+ N_2$ magnetic (anti)quarks and $(\tilde{N}_f)^2$ mesons. If we gauge
the $U(N_2)$ subgroup of the $U(N_f + N_2)$ flavor group we obtain
exactly the matter content of the electric and magnetic theories of
our example. Actually the fields $M$, $X$, $r$ and $\tilde r$ are just
the result of the decomposition of the original mesons and therefore
there exist several relations between the $\RR$ charges of the fields.
\baq & \RR_i(M) = 2 \RR_i(Q) ~, \RR_i(X) = 2 \RR_i(N) ~,
   \RR_i(r) = \RR_i(Q) + \RR_i(N) ~, & \\
 & \RR_i(q) = 1 - \RR_i(Q) ~, \RR_i(n) = 1 - \RR_i(N) ~, i=1,2 ~, &
\eaq
where the first line follows from the identification of the
mesonic operators in the electric and magnetic theory \equ{mesons},
the second line follows from the first line and the fact that the
superpotential \equ{superpot} has $\RR$ charge $2$.

\section{Can brane configurations be related to
non-- supersymmetric gauge theories and dualities ?}

A natural question to ask after passing from theories with $N=2$ to
those with $N=1$ is what are the  brane setups that describe
non-supersymmetric 4D gauge theories and in particular models with
Seiberg like dualitites.

A priori it is not clear whether non-supersymmetric brane configurations
represent consistent stable backgrounds.
The stability of supersymmetric brane configurations is based on
the no-force situation of static multiple brane solutions of the
low-energy effective action of type IIA and type IIB string theories
\cite{BREJS,AEH,AVVV}.
This is lost when supersymmetry is completely broken by the brane background
configuration; the BPS condition is then no longer satisfied.

We thus have to expect that the brane configurations we are about to
discuss are {\it unstable}. Note that also the stability of the 
configurations considered
in \cite{egk} should be carefully re-examined,
 as there is a unbalanced force on the
NS branes from the D branes stretched between them \cite{witten}.

In what follows we will ignore this apparent problem and consider
these configurations. We will apply the same moves that were used in
the supersymmetric cases and compare the matter content looking for
cases in which 't Hooft anomaly matching conditions are satisfied. In
those cases in which they are satisfied we may have candidates for
dual pairs. (Thus, if at all, the justification for looking at these
configurations is a posteriori.)

In the following table we list several brane configurations which
break all supersymmetries. The table includes only branes
which are along some coordinate axis and not branes at angles.
We denote the NS$^{(1)}$, NS$^{(2)}$ and D6 branes by their
world volume coordinates which they have in addition to $x^0,x^1,x^2, x^3$.
The table also includes possible $U(1)$ symmetries that have their
origin in a rotation invariance in a subspace of 10-d spacetime.
\be
\begin{array}{lcccccc}
             &{\rm NS}^{(1)} & {\rm NS}^{(2)} & {\rm D6} &   U(1) &
{\rm D4}\ {\rm created}  & \\[2mm]
(i)            &   x^4,x^5        &
  x^4,x^7        & -         & (89)      & - \\
(ii)   &
  x^4,x^5    & x^4,x^5          & x^4,x^8, x^9  &  (89)   & {\rm no} \\
(iii)      &  x^4,x^5    &  x^8, x^9         & x^4,x^5,x^8            &
(45)      & {\rm no}     \\
(iv)      &  x^4,x^5    &  x^8, x^9         & x^4,x^7,x^8            &
{\rm no}      & {\rm no}     \\
(v)      &  x^4,x^5    &    x^4,x^7       & x^7,x^8,x^9            &
(89)      & {\rm yes}    \\
(vi)      &  x^4,x^5    &    x^4,x^7       & x^4,x^5,x^7            &
(89)      & {\rm no}    \\
(vii)      &  x^4,x^5    &    x^4,x^7       & x^5,x^7,x^8            &
{\rm no}      & {\rm no}    \\

\end{array}
\eel{qunr1}\\[2mm]
Each item describes one example out of a family of similar  configurations.
For instance NS$^{(2)}$ in $(i)$ can be any of  the following possibilities.
$x^i, x^j$ with $i\in\lbrace 4,5\rbrace$ and $j\in\lbrace 7,8,9\rbrace$.

It is  now straightforward to realize that
indeed all the supersymmetries are broken for the setups of the table.
Let us demonstrate it for the cases that NS$^{(2)}$ is  $x^4,x^7$.
The supersymmetry parameters have to obey
\be
\epsilon_L = \Gamma^0 \cdots \Gamma^5\epsilon_L; \qquad
\epsilon_R = \Gamma^0 \cdots \Gamma^5\epsilon_R
\label{mish01}\ee
due to the NS brane and
\be
\epsilon_L = \Gamma^0 \cdots \Gamma^4\Gamma^7\epsilon_L; \qquad
\epsilon_R = \Gamma^0 \cdots \Gamma^4\Gamma^7\epsilon_R
\label{mish02}
\ee
due to the \nss brane.
Substitute $\epsilon_L$ from \equ{mish01} into \equ{mish02} to get
\be
\epsilon_L=-
\Gamma^5\Gamma^7\epsilon_L=\Gamma^5\Gamma^7\Gamma^5\Gamma^7
\epsilon_L=-\epsilon_L
\ee
which obviously means that there is no non-trivial solution for $\epsilon_L$
and similarly for $\epsilon_R$.

Seiberg's duality followed in the supersymmetric models from brane
configurations that have D6 branes that, when moving them along the $x^6$
direction past the NS5 brane, necessarily meet the NS5 brane in
space-time. For this it is necessary that the NS and the D6 branes
have no common transverse direction except $x^6$.
Only then do we create new D4 branes.
The last column of the table indicates whether the model
has this property.

\subsection {Seiberg's duality in $N=0$}
\begin{figure}[ht]
\hbox to\hsize{\hss
\epsfysize=4cm
\epsffile{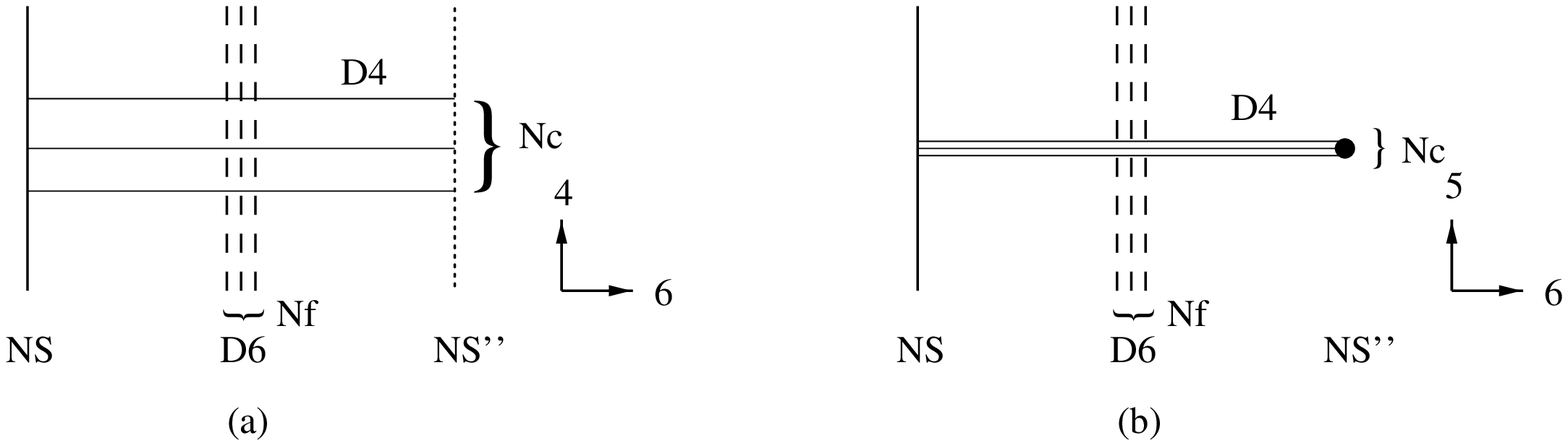}\hss}
\vspace{0cm}
\caption{} \label{pic8}
\end{figure}

As indicated in the table model $(v)$
can incorporate non-trivial NS brane rearrangements.
The basic setup for this case, which is described in
figure \ref{pic8}, is related to the configuration of
Figure 1, with an NS brane along $ x^4,x^7$,
denoted by \nss, replacing the left side NS.

We want to emphazise again the issue of the stability of the
brane configuration. Being aware of potential problems, we nevertheless
proceed in the same way as we did when considering supersymmetric
configurations.

The fermionic content of the field theory
is summarized in the following table
\be
\begin{array}{lcccccc}
             &~~U(N_c) & SU(N_f) & SU(N_f) &   U(1)_B &
U(1)_R & \\[2mm]
\psi_Q            &   N_c        &
N_f        & 1         & {1\over N_c}      & {-N_c\over N_f} \\
\psi_{\tilde{Q}}   & \bar
N_c     & 1          & \bar N_f  & {-1\over N_c}      & {-N_c\over N_f} \\
(\lambda/\psi)_{el}      &  N_c^2     & 1          & 1         & 0      &
1     \\
\end{array}
\eel{qunr}\\[2mm]
where  the masless fermions were classified according
to the maximal global symmetry given
in the first line. $U(1)_R$ is the axial symmetry unbroken by instantons.
$\psi_Q$ and $\psi_{\tilde Q}$, the ordinary quarks and anti-quarks,
are associated with the 4-6 strings, and
$\lambda_{el}$ are the ``gauginos".

Actually there are two adjoint fermions,
the gauginos  that originate from $W_\alpha$
and a second  adjoint fermion $\psi_{adj}$
that is associated with $\Phi$. Note that in the present model
there is a real adjoint scalar field related
to the motion of the D4 along $x^4$ whereas its $N=2$ companion,
the second real scalar field related
to the motion of the D4 along $x^5$, is absent.
The reason that we have included only one adjoint fermion is that
there is only one $U(1)_R$ symmetry unbroken by instantons,
since the charges of $\lambda$ and $\psi$ are not independent due to a Yukawa
term. Therefore, there is only one linear combination of the fermions, denoted
by $(\lambda/\psi)_{el}$ that remains massless.
Note also that in non-supersymmetic field theories chiral symmetry
protects the masslessness of fermions and not of their associated ``scalar
partners".

Performing a sequence of moves that is identical to the one performed for the
$N=1$ case we end up with a brane configuration similar to the one described in
Figure 4. where again \nss replaces \ns.
Using the same rules  to convert the branes and  the strings between them to
fields on the 4D world-volume as for the electric theory one now finds the
following fields
\be
\begin{array}{lcccccc}
             &~~U(N_f - N_c) & SU(N_f) & SU(N_f) &   U(1)_B &
U(1)_R & \\[2mm]
 \psi_q            & N_f -  N_c      & \overline{N_f}
&   1  & {1\over N_f-N_c} &
{N_c\over N_f}-1\\
\psi_{ \tilde{q}}    &\overline{N_f - N_c}     & 1   &  N_f  &
-{1\over N_f-N_c}&{N_c\over N_f}-1\\
 \lambda_{mag}    &  {(N_f -N_c)}^2     & 1          &
1         & 0     &1  \\
 \psi_M           &        1           & N_f        & \bar N_f &
0      &-2{N_c\over N_f}+1    \\

\end{array}
\eel{qunr2}\\[2mm]

The emergence of the magnetic quarks, anti-quarks and  ``magnetic
gaugino"
 from strings is similar to that in
$N=1$ theory.
The mesons, as in the $N=1$ cases, are again due to
strings among the $N_f$ D4 branes streched between the
D6 branes and the \nss branes. However, now
the   D4 brane  can move  only along one direction, $x^7$.
Following our dictionary this implies real and not complex
$U(N_f)$ real scalar fields. In terms of the field theory
mesons are composites of $\phi_M\sim \psi_Q\psi_{\tilde Q}$
whereas the mesonic fermions
$\psi_M$ are composites of
$\psi_M\sim \psi_Q\lambda\psi_{\bar Q}$ with
$U(1)_R$ charge $-2{N_c\over N_f}+1$.
The masslessness of these fermions is protected  by  $U(1)_R$ symmetry.
In fact without any additional  term in its action  the magnetic theory has
an  additional $U(1)$ symmetry which does not have a counterpartn in the
electric one. This is obviously avoided
by the introduction of a Yukawa  term
which is the remnant of the
$MQ\tilde Q$ superpotential term in the $N=1$ magnetic theory.

It is now straightforward to check that `t Hooft anomalies associated with

$SU(N_f)^3,~SU(N_f)^2\times U_B(1),~SU(N_f)^2U(1)_R,~U(1)_R$

$(U(1)_R)^3,~(U(1)_R)^2 U(1)_B,~(U(1)_R)^2 U(1)_B,~U(1)^3_B$

\noindent
of the electric and magnetic theories match.

In section 3.2  certain moves of brane configurations led to a duality
between $U(N_c)$ and $U(kN_f-N_c)$ gauge theories where $k$ was the
number of NS branes taken instead of the basic configuration with one NS
brane.
Repeating the same structure with \nss replacing the \ns brane establishes a
similar non-supersymmetric duality.

The magnetic theory includes now a set of $k$ composit mesons
$\psi_M\sim \psi_Q(\lambda)^{2l-1}\psi_{\bar Q}$ with $l=1,\dots,k$.
Their $U(1)_R$ charges are
$-2{N_c\over N_f}+2l-1$. The sum of the $U(1)_R$ charges and the sum of
their cubes are $-N_c^2-1$ and $-{2N_c^4\over N_f^2} +N_c^2-1$, respectively
and are identical to the corresponding anomaly factors in the electric
theory.

Several remarks are now in order:

(i) The original supersymmetric duality was supported by the following
tests: 't\ Hooft anomaly equations, chiral rings, mass and vev deformations.
In the non-super\-sym\-metric analogs only the first test is available.
Moreover, recall that 't Hooft anomaly matching conditions are only necessary
conditions to maintain chiral symmetry but by no means sufficient ones.

(ii) The $N=1$ magnetic theory is characterized by a moduli space
parametrized by expectation values
$\langle q\rangle, \langle\tilde q\rangle$ and $\langle M\rangle$. The
non-supersymmetric theory, where the dual squarks and mesons are massive,
has a unique minimum of the potential at the origin of the moduli space
$\langle q\rangle =\langle\tilde q\rangle=\langle M\rangle=0$ \cite{APSY}.

(iii) Seiberg's duality was studied in the context of softly broken
supersymmetric QCD  in \cite{APSY}. It was shown there that indeed for the
case  that $R$ is not broken and the ``gaugino" remains massless
one can justify  the duality, at least for small squark masses.
However, QCD inequalities seem to forbid massless fermionic mesons
in the decoupling limit of large squark masses.
In the case of no massless adjoint fermion one can show that there are no
non-trivial solution to `t Hooft anomaly equations and thus there is no
such duality. Notice, that the model (v) has, at least classically
a $ U(1)_R$ symmetry.

(iv) The setup with NS and \nss can be acheived from that
with NS and NS$^\prime$
by performing a complex rotation of $z\rightarrow e^{i\theta}z\ \
w\rightarrow e^{i\theta}w$ with $z=x^8+ix^4$ and  $w=x^9+ix^7$ similar to the
one suggested in \cite{barbon}. The construction of explicit soft susy
breaking terms is currently under investigation.

(v) As we have emphasized at the beginning of this section it is difficult
to see how the non-supersymmetric configurations could be consistent string
backgrounds. If indeed they are unstable and one can argue that any gauge
theory that admits Seiberg's duality has to have a brane counterpart,
this may indicate that the $ U(1)_R$ of the $R$ model discussed in \cite{APSY}
is broken  for any squark mass.
Conversly, if one finds  that there is a region where $ U(1)_R$ is unbroken,
it might be a hint that there should be a stable brane configuration
describing it.

\section{Instantons and brane configurations}

The brane configurations described in the previous sections correspond to
the perturbative regime of certain supersymmetric and non-supersymmetric
4D gauge  field theories. We now want to
explore the possibility of translating
non-perturbative gauge fields, namely instantons, into brane
configurations.
In \cite{han} Euclidean D1 branes were introduced in the type IIB
theory to  describe 3D instantons.
This was further explored in ref. \cite{boer1,boer2}.

The branes that we have in our arsenal to induce 4D instanton effects
are the followig objects in their Euclidean formulation:
even branes of IIA and the NS5 branes.

Let us indentify now the conditions one has to
impose on type IIA branes describing 4D instantons.\footnote{We 
thank O. Aharony for  pointing out to us 
the identification of the D0 branes with zero -size instantons.}

(i) To describe a point in the 4D world volume of the D4 branes,
the directions $x^0,x^1,x^2,x^3$ have to be
transverse directions to the brane. 

(ii) Since in  4D  gauge theories instantons
break ${1\over 2}$ of the supersymmetries,
the additional   brane should break
 the $N=2$ theory down  to $N=1$, and  break supersymmetry all together in
the $N=1$ theory.

(iii) The branes added should ``communicate" with the D4 branes.
In general this can be achived if they  end on the D4 branes, or
if there are open strings streched between them and
the D4 branes.

(iv) The action of the instantons should be finite and be proportional
to ${1\over g^2}$.

(v) There should not be any (non-singular) 
 instanton effect in the brane setup  that
corresponds to the  abelian theory.

{}From (i) it follows that one can add Euclidean D0, D2, D4 branes
 with world volumes embedded in the
$\lbrace x^4,x^5, x^6, x^7, x^8, x^9\rbrace$ subspace of ${\bf R}^{1,9}$.
If 
we demand that the new branes will end on the NS 5branes or the 
D4 branes,  that will leave
only D0 and D2 branes and Euclidean NS strings.
Checking the Euclidean D2 branes one finds that the branes along
$x^4, x^5, x^6$ and  $x^6, x^8, x^9$ break supersymmetry altogether
in the $N=2$ case whereas the branes with
$x^4, x^6, x^i$ (or $x^5, x^6, x^i$) where $i\in\lbrace 7,8,9\rbrace$
have the required supersymmetry breaking pattern. Recall that an Euclidean
D brane imposes  a condition on the parameter of supersymmetry transforamtion
that differs by a factor of $i$ for the one imposed by an ordinary Minkowski
brane. For instance the $x^4, x^5, x^6$ brane requires that
$\epsilon_L = i\Gamma^4\Gamma^5\Gamma^6\epsilon_R$.
The problem with the Euclidean D2 branes is that their action, which is
proportional to their volume, always involves one non-compact direction and
thus is infinite.

We are left with the D0 branes streched along $x^6$ between the NS
branes. Their action is  proportional to 
$| x^6_{{\rm NS}_1} - x^6_{{\rm NS}_2}|=
\lambda_{string}/g^2$, with $\lambda_{string}$ being the string coupling
constant. Since  there is an extra 
${1\over\lambda_{string}}$ factor  associated with any D brane action,  
the D0 action  is  proprotional to $1/g^2$.
The D0 branes affect the D4 branes via the open strings that are streched
between them. At first sight they seem to violate condition (v) since they 
can exist also in case of   single D4 brane.  In fact our claim 
is that they correspond to zero-size instantons which  can show up also 
in the abelian theory.
This agrees also with arguments given in \cite{horwit} which identify 
D(p-4) branes in Dp branes as zero size instantons. 
Finite size instantons are associated with the Higgs
phase on their world line. The brane picture of the
phase transition from  zero to finite size instantons, as well as
questions like how is the $U(1)_R^{45}$ symmetry broken by
fermionic zero modes
 are under current
investigation.      
It is interesting to note,
 that the D0 branes  
can be obtained from the 3D type IIB setup
by compactifying $x^3$ on a circle on which D1 brane is wrapped and
performing T-duality along $x^3$. 
This is the same duality needed
to pass from the IIB brane configurations associated with the 3D physics 
to the type IIA brane configurations of \cite{egk}.

\section{Discussion}

String backgrounds that include brane configurations
can be related to the 4D field theory that describes the low energy
phase of nature in the following  ways:
(i) The universe may be assciated with
a D3 brane ( or a truncated D4 ) which is part of the string
background. (ii)  Physics on branes may shed certain new light on
field theories even without the scenario mentioned in (i).
The recent progress in the interplay between brane physics and
field theory that originated in  \cite{han} belongs to (ii).
It should be interesting to explore also option (i).

In the present paper we have unraveled an additional
piece  in the puzzle of the full relations
between field and brane theories.
We elaborate on the brane rearrangements that manifest
the equivalence between the Higgs branches of
``dual" $N=2$ models. We provided further
evidence for the construction of \cite{egk} by analyzing
the correspondence between the
mass and vev deformations
of the $N=1$ electric and magnetic theories.
This was done for the setup that translates to
the original  duality of Seiberg\cite{sei1}
as well as to the model that includes
adjoint chiral multiplet\cite{kut}.
We also suggested a new construction that corresponds to
a Seiberg duality in field theories with product gauge
groups.
A great  challange in the inteplay between string backgrounds
and 4D field  theory is the case without supersymmtry.
Unfortunately, we were not able to suggest a mechanism that
stabilizes the brane constructions we have considered.
However, by closing our eyes to this problem we  discussed a
model that may be related to a duality of softly broken
supersymemtic QCD model with unbroken $U(1)_R$ symmetry.
Lastly, we enlist certain conditions that we found to be
plausible in the search for branes that may mimic gauge
instantons. 
We have elaborated on the possible identification of the instantons within
the brane picture. 
It seems to us that
D0 branes embedded in a D4 brane  may correspond to zero size instanton. 

Certainly  the journey toward  a full understanding of the inteplay
between brane configurations of type II string theories and gauge
field theories in 3D and 4D is still in its infancy and
there are many open questions to address. Let us mention several
of them related to the present work:

(i) The constructions of \cite{han} and \cite{egk} will be proven
to be more powerful if novel dualities are discovered.
Generalizing  the way we have built the model with product groups
should lead to novel field theory  dual pairs.

(ii) A proper understanding of brane instantons might enable us to
explore, directly in the brane picture,  
field theoretic instanton phenomena such as their
contribution to the complex coupling $\tau$ in $N=2$ theories,
the superpotential of $N=1$ theories with $N_C>N_f$,
the breaking of the axial $U(1)$ ( $U(1)_{45}$) etc.
In fact, by considering other possible brane attachements not
obeying the conditions of section 6, one may identify additional
non-perturbative field theory phenomena.
This immediately raises the more general question of the relation
between the space of 4D field theories and the space of
consisitent brane configurations.

(iii) In this context one should also raise the question of stability
of the brane configurations and whether it is necessary for
deriving field theory results.

(iv) Recently, the brane setup for $N=2$ was analyzed,  yielding
explicit solutions for the
Coulomb branch of a large family of four-dimensional
N=2 field theories with zero,
negative or positive  beta function\cite{witten}.
There are numerous
questions that one can address using those new results in relation
to the topics discussed in our work.

{\bf Acknowledgements}
We  would like to thank O.\ Aharony, D.\ Ghoshal and D.\ Kutasov
for useful conversations.

\end{document}